\documentclass[aps,prl,twocolumn,amsmath,amssymb,floatfix,nofootinbib,superscriptaddress]{revtex4-1}

\newcommand{\nhat}{\hat{ \mathbf{n}}}

\usepackage{graphicx}
\usepackage{dcolumn}
\usepackage{bm}
\usepackage[usenames, dvipsnames]{color}
\usepackage[normalem]{ulem}
\usepackage{xcolor}

\graphicspath{ {figures/} }

\usepackage{epstopdf}
\newcommand{\be}{\begin{eqnarray}}
\newcommand{\non}{\nonumber \\}
\newcommand{\ee}{\end{eqnarray}}

\newcommand{\vpot}{\Upsilon}

\usepackage[export]{adjustbox}

\def\vl{\boldsymbol{\ell}}
\def\vlp{\boldsymbol{\ell}'}

\def\nhat{\hat{\mathbf{n}}}

\newcommand{\aap}{Astron. Astrophys.}

\newcommand{\mnras}{Mon. Not. R. Astron. Soc.}
\newcommand{\jcap}{J. Cosm. Astropart. Phys.}

\newcommand{\dd}{{\rm d}}

\begin{document}

\title{Transverse Velocities with the Moving Lens Effect}
\newcommand{\cita}{Canadian Institute for Theoretical Astrophysics, University of Toronto, 60 St.~George Street, Toronto, ON M5S 3H8, Canada}
\newcommand{\imperial}{Astrophysics Group \& Imperial Centre for Inference and Cosmology, Department of Physics, Imperial College London, Blackett Laboratory, Prince Consort Road, London SW7 2AZ, UK}
\newcommand{\perimeter}{Perimeter Institute for Theoretical Physics, 31 Caroline St N, Waterloo, ON N2L 2Y5, Canada}
\newcommand{\york}{Department of Physics and Astronomy, York University, Toronto, ON M3J 1P3, Canada}
\newcommand{\smu}{Department of Physics,
Southern Methodist University, 3215 Daniel Ave, Dallas, TX 75275, U.S.A.}

\author{Selim~C.~Hotinli}
\affiliation{\imperial}

\author{Joel~Meyers}
\affiliation{\smu}

\author{Neal~Dalal}
\affiliation{\perimeter}

\author{Andrew~H.~Jaffe}
\affiliation{\imperial}

\author{Matthew~C.~Johnson}
\affiliation{\york}
\affiliation{\perimeter}

\author{James~B.~Mertens}
\affiliation{\york}
\affiliation{\perimeter}

\author{Moritz~M\"{u}nchmeyer}
\affiliation{\perimeter}

\author{Kendrick~M.~Smith}
\affiliation{\perimeter}

\author{Alexander~van~Engelen}
\affiliation{\cita}

\date{\today}


\begin{abstract}

Gravitational potentials which change in time induce fluctuations in the observed cosmic microwave background (CMB) temperature.  Cosmological structure moving transverse to our line of sight provides a specific example known as the moving lens effect.  Here we explore how the observed CMB temperature fluctuations combined with the observed matter over-density can be used to infer the transverse velocity of cosmological structure on large scales.  We show that near-future CMB surveys and galaxy surveys will have the statistical power to make a first detection of the moving lens effect, and we discuss applications for the reconstructed transverse velocity.

\end{abstract}

\maketitle


\paragraph{\textbf{Introduction}}
Upcoming surveys of the cosmic microwave background (CMB) including those by 
Simons Observatory~\citep{Ade:2018sbj} and CMB-S4~\citep{Abazajian:2016yjj} and galaxy surveys such as the Dark Energy Survey (DES)~\citep{2005astro.ph.10346T} and the survey by the Large Synoptic Survey Telescope~(LSST)~\citep{Abell:2009aa}, will provide new opportunities for novel cosmological measurements.  In particular, by using the CMB as a cosmological backlight, secondary fluctuations induced by the interaction of CMB photons with structure along the line of sight allow for new methods to study the history and evolution of the Universe. Such second-order effects include weak gravitational lensing by large-scale structure (see~\citep{Lewis:2006fu} for a review); the integrated Sachs-Wolfe (ISW)~\citep{1967ApJ...147...73S} and Rees-Sciama effects~\citep{1968Natur.217..511R}, describing the process by which time-dependent gravitational potentials alter the energy of CMB photons; and the Sunyaev-Zel'dovich (SZ) effect~\citep{Zeldovich:1969ff,Sunyaev:1970er,Sunyaev:1972eq,Sunyaev:1980vz,Sazonov:1999zp}, whereby CMB photons undergo Compton scattering with free electrons in galaxy clusters and the intergalactic medium.

Here we focus on the moving lens effect~\citep{1983Natur.302..315B} as a source of secondary CMB anisotropies and estimate the prospects for detecting the effect with upcoming observations.  The temperature fluctuations imprinted by the transverse motion of individual objects are expected to be weak and can be easily confused with other effects, which makes detection challenging~\citep{Tuluie:1995ut,Aghanim:1998ux,Cooray:2002ee}. 
We consider a new statistical approach to detecting the moving lens effect, which effectively combines the signal from the many objects with a common bulk motion. Using this approach, we demonstrate that data expected from upcoming CMB experiments and galaxy surveys should have the statistical power to make a detection of the moving lens effect at high significance.

\begin{figure}[t!]
    \centering
    \includegraphics[width = \columnwidth]{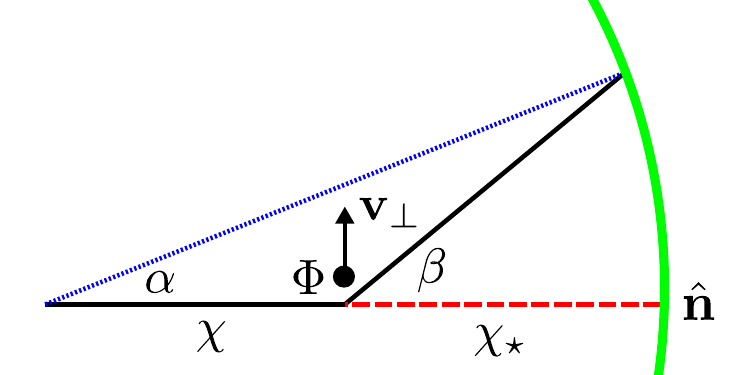}
    \caption{Sketch of the geometry in the CMB rest frame, for a lens of potential $\Phi$ moving with transverse velocity $\mathbf{v}_\perp$, as seen by an observer at comoving distance $\chi$ from the lens, and distance $\chi_\star$ from the CMB last scattering surface.
}
    \label{fig:sketch}

\end{figure}

A gravitational potential moving with velocity $\mathbf{v}_\perp$ transverse to our line-of-sight direction $\nhat$ leads to CMB temperature fluctuations given (at lowest order) by 
\be\label{eq:T_mv}
    \Theta (\nhat) = \mathbf{v}_\perp \cdot \boldsymbol{\beta}(\chi \nhat)\ , 
\ee
where $\Theta = \Delta T / T$ is the fractional CMB temperature fluctuation, $\chi$ is the conformal distance, and  $\boldsymbol{\beta}$ is the deflection angle as seen by the lens~\citep{1983Natur.302..315B,1998A&A...334..409A,Gurvitz:1986ab,2002PhRvD..65h3518C,Lewis:2006fu}; see Fig.~\ref{fig:sketch}.  We can understand the origin of this effect in a few physically equivalent ways.

The motion of an observer with respect to the CMB induces a kinematic dipole temperature anisotropy due to the Doppler boosting of the CMB monopole, and also results in angular aberration of CMB fluctuations~\citep{Challinor:2002zh,Aghanim:2013suk}.  We define the CMB rest frame as the reference frame in which the aberration of the CMB fluctuations vanishes, which is not identical to a frame in which the temperature dipole vanishes.  The observed temperature dipole in the rest frame of the Solar System has an amplitude of about $10^{-3}$~\citep{Aghanim:2018fcm}, while the anticipated intrinsic component (the amplitude in the CMB rest frame) is on the order $10^{-5}$, and so the CMB rest frame is often approximated by boosting to a frame in which the observed dipole vanishes~\citep{Notari:2011sb}.  

In the rest frame of the CMB, a massive object moving transverse to the line of sight of a stationary observer generates a gravitational potential which evolves in time.  As CMB photons traverse this time-dependent potential, they receive a redshift or blueshift in close analogy with the ISW effect   $\Theta (\nhat) = -2 \int_0^{\chi_\star}  \mathop{\dd\chi} \mathbf{v}_\perp \cdot \boldsymbol{\nabla}_\perp \Phi(\chi \nhat)=\mathbf{v}_\perp \cdot \boldsymbol{\beta}(\chi \nhat)$, where $\chi_\star$ is the conformal distance to the surface of last scattering and $\Phi$ is the gravitational potential.  This induces a characteristic dipole pattern of CMB temperature fluctuations oriented along the object's transverse velocity.

Next, viewed from the rest frame of the lens, this effect can be recast as lensing of the (kinematic) CMB dipole seen by the lens.  The photons deflected toward the observer have a temperature $T(1+\mathbf{v}_\perp
\cdot(\nhat+\boldsymbol{\beta}))$, giving at lowest order $\Theta(\nhat) = \mathbf{v}_\perp \cdot \boldsymbol{\beta}$ after transforming to the observer frame.

Finally, the calculation for an observer moving with the same peculiar velocity as the lens with respect to the CMB is slightly more subtle. 
Photons deflected into the line of sight of the observer by gravitational lensing originate from the surface of last scattering separated from the observation direction $\nhat$ by an angle $\boldsymbol{\alpha}$.
In this moving frame, the CMB temperature has a kinematic dipole of the form $T_0[1+\mathbf{v}\cdot\nhat]$. In the standard treatment, lensing remaps the observed temperature according to $T(\nhat) =\tilde{T}(\nhat + \boldsymbol{\alpha}) = \tilde{T}(\nhat) + \boldsymbol{\nabla} \tilde{T}(\nhat) \cdot \boldsymbol{\alpha}(\nhat) + \cdots$ where $\tilde{T}$ is the unlensed temperature, and in this case gives $\Theta = \mathbf{v}_\perp \cdot \boldsymbol{\alpha}$ at lowest order. It is clear that this differs from what was calculated above since $\boldsymbol{\alpha} \neq \boldsymbol{\beta}$.  However, one must be careful to take into account the fact that the photons which are deflected into the line of sight of the observer were not emitted perpendicular to the surface of last scattering (an effect which is formally of the same order as the lensing deflection).  This change to the emission angle is usually negligible for CMB temperature fluctuations~\citep{Lewis:2017ans}, but it cannot be ignored in this case since the dominant temperature source at the surface of last scattering is due to the Doppler effect and therefore has an intrinsic dipole anisotropy.  The emission angle relative to the line of the sight to the lens is $\boldsymbol{\beta}$, and so the observed temperature fluctuation evaluated in the frame comoving with the lens is $\Theta(\nhat) = \mathbf{v}_\perp \cdot \boldsymbol{\beta}(\chi\nhat)$.  It has previously been shown that one can also arrive at this expression by treating the kinematic component of the dipole as a source at infinite distance~\citep{Cooray:2005my}.  This analysis also demonstrates that the CMB dipole measured in the rest frame of the CMB (the intrinsic dipole) is physically distinct from the dipole induced by boosts away from that frame (the kinematic dipole)~\citep{Lewis:2006fu}, and the former can therefore be reconstructed by measuring how it is lensed~\citep{Meerburg:2017xga}.


\paragraph{\textbf{Estimator}}     

We wish to construct a quadratic estimator for the transverse velocity field $\mathbf{v}_\perp(\nhat,z)$ on large angular scales ($\ell\lesssim100$), given maps of the CMB temperature and of a tracer of the density field at some redshift on small angular scales ($\ell\gtrsim2000$), 
analogous to a CMB lensing quadratic estimator~\citep{Hu:2001kj}, for example.  Our focus is on the large-scale velocity field, where we anticipate that the velocity is linear and curl-free, such that we can define a transverse velocity potential $\vpot(\nhat,z)$, with $\mathbf{v}_\perp(\nhat, z) = \boldsymbol{\nabla}{\vpot}(\nhat, z)$.
We utilize the typical definition of the gravitational lensing potential $\phi$ such that $\boldsymbol{\alpha} = \boldsymbol{\nabla}\phi$, with
\be
    \phi(\nhat) = -2 \int_0^{\chi_\star} \mathop{\dd \chi} \frac{\chi_\star - \chi}{\chi_\star \chi} \Phi(\chi\nhat) \, ,
\ee
where we have assumed spatial flatness.  We can construct a similar potential for the deflection as seen by the lens
\be\label{eq:psi_field}
    \psi(\nhat) = -2 \int_0^{\chi_\star} \mathop{\dd \chi} \frac{1}{\chi} \Phi(\chi\nhat) \, ,
\ee 
such that $\boldsymbol{\beta} = \boldsymbol{\nabla}\psi$, and which differs from the ordinary lensing potential $\phi$ by a ratio of the lens and source distances.

Given  an observed map of the CMB temperature, $\Theta^\mathrm{obs}$, and a map of $\psi^\mathrm{obs}$ as derived from, for example, a survey of large-scale structure, we can write the desired quadratic estimator as
\be\label{eq:esimator}
    \hat{\vpot} (\mathbf{L}) = N(\mathbf{L}) \int \frac{\dd^2 \boldsymbol{\ell}}{(2\pi)^2} g(\boldsymbol{\ell},\mathbf{L}) {\Theta^\mathrm{obs}(\boldsymbol{\ell})} \psi^\mathrm{obs}(\mathbf{L}-\boldsymbol{\ell}) \, .
\ee
We have suppressed the redshift dependence of $\hat{\vpot}$ and $\psi$, and the normalization $N(\mathbf{L})$ and filter  $g(\boldsymbol{\ell},\mathbf{L})$ are to be determined.   We are using the flat-sky approximation so that $\boldsymbol{\ell}$ and $\mathbf{L}$ are two-dimensional Fourier wavevectors, and have found the results agree well with a full-sky estimator, as is also the case with lensing estimators~\citep{Okamoto:2003zw}. 
Following, e.g., Ref.~\cite{Hu:2001kj}, we minimize the estimator variance subject to the constraint that the estimator is unbiased, i.e., that $ \vpot(\mathbf{L}) = \left\langle \hat{\vpot}(\mathbf{L}) \right\rangle\! _{\Theta,\psi}$. 
At lowest order, the variance is
\begin{align}
    &\left\langle  \hat{\vpot}(\mathbf{L})\hat{\vpot}(\mathbf{L}')  \right\rangle = (2\pi)^2 \delta^{(2)}(\mathbf{L}+\mathbf{L}') \Big[ C_L^{\vpot\vpot}  + N(\mathbf{L})\Big]\, , 
\label{eq:v_variance}
\end{align}
where the transverse velocity potential power spectrum is defined as
\be\label{eq:tr_signal}
    C_\ell^{\vpot\vpot} = \frac{4\pi}{\Delta \chi} \int_{\chi_\mathrm{min}}^{\chi_\mathrm{max}} \dd \chi \int \frac{\dd k}{k} \frac{\mathcal{P}_v(k,\chi)}{(k\chi)^2} \left[j_\ell(k\chi)\right]^2 \, ,
\ee
and $\mathcal{P}_v$ is the dimensionless power spectrum of the three-dimensional velocity $|\mathbf{v}|$.
We find that we must fix the normalization to
\be
    N(\boldsymbol{L}) = \left[\int \frac{\dd^2 \boldsymbol{\ell}}{(2\pi)^2} 
    C_{|\boldsymbol{\ell} -\mathbf{L}|}^{\psi\psi} g(\boldsymbol{\ell},\mathbf{L}) 
    \mathbf{L} \cdot (\mathbf{L}-\boldsymbol{\ell})\right]^{-1} \, ,
\ee
and that the filter which minimizes the variance is 
\be
    g(\boldsymbol{\ell}, \mathbf{L}) =  \frac{\mathbf{L} \cdot (\mathbf{L}-\boldsymbol{\ell})}{C_\ell^{\Theta\Theta,\mathrm{obs}}} \frac{C_{|\boldsymbol{\ell} -\mathbf{L}|}^{\psi\psi}}{C_{|\boldsymbol{\ell} -\mathbf{L}|}^{\psi\psi,\mathrm{obs}}} \, ,
\ee
thereby giving for the noise on a reconstructed mode
\be\label{eq:noise_naive}
    N(\boldsymbol{L},z)\!=\!\left[\!\int \frac{\dd^2 \boldsymbol{\ell}}{(2\pi)^2} 
    \frac{\left[\mathbf{L} \cdot (\mathbf{L}-\boldsymbol{\ell})\right]^2}{C_\ell^{\Theta\Theta,\mathrm{obs}}} \frac{\left(C_{|\boldsymbol{\ell} -\mathbf{L}|}^{\psi\psi(z)}\right)^2}{C_{|\boldsymbol{\ell} -\mathbf{L}|}^{\psi\psi(z),\mathrm{obs}}} 
    \right]^{-1}\!\! 
\ee
where we reintroduced the redshift dependence of our noise estimate. 

In the above derivation, we ignored other secondary CMB fluctuations which may contribute to the estimator in Eq.~\eqref{eq:esimator}.  We discuss such biases and their mitigation after making an initial estimate of the signal-to-noise ratio for our estimator of the moving lens effect.


\begin{figure}[t!]
    \centering
    \includegraphics[width = \columnwidth]{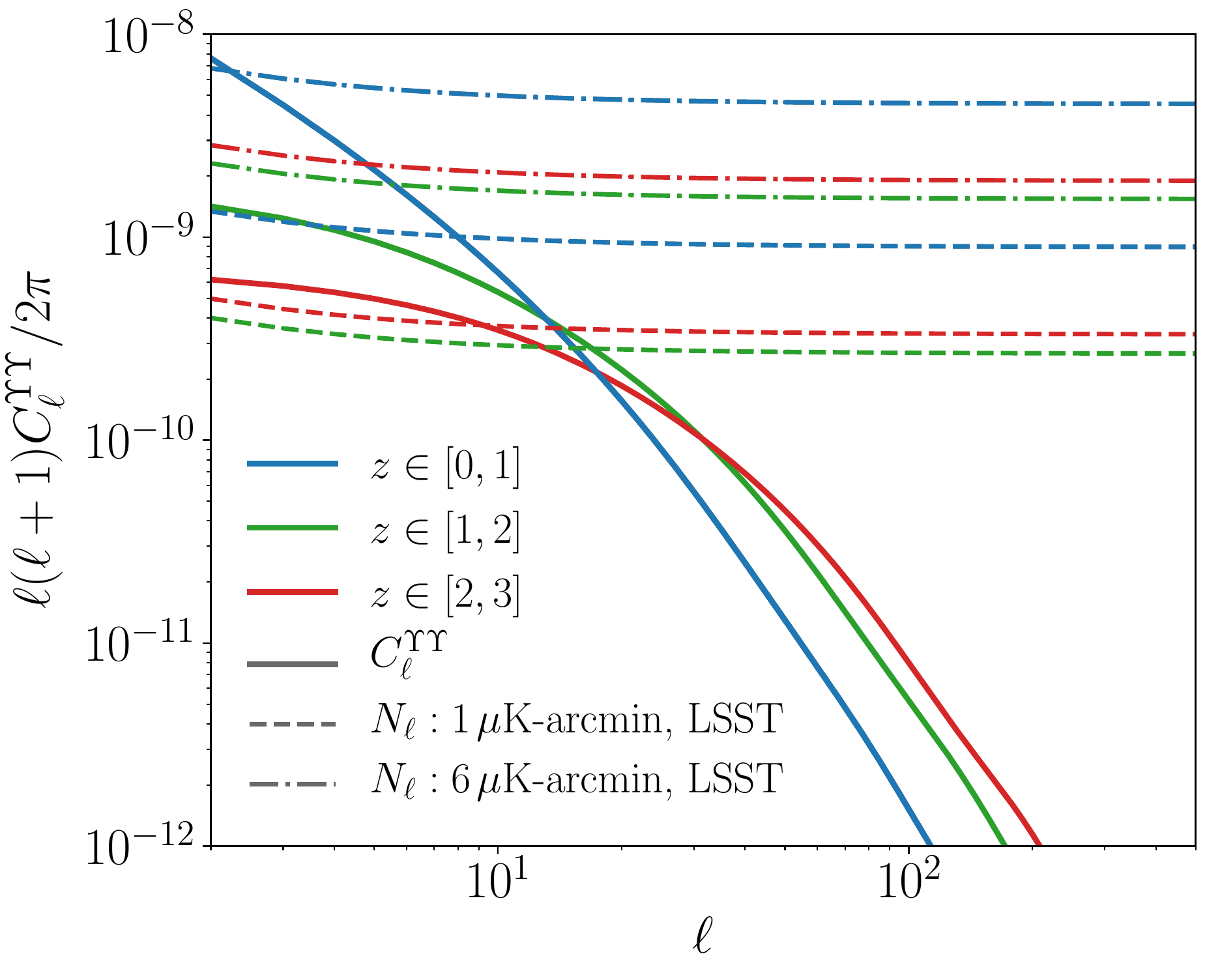}
    \caption{Power spectrum of the transverse velocity potential and reconstruction noise in several redshift bins for two CMB experiments with a 1.4-arcmin beam combined with LSST. Where the signal  curves, in solid, exceed the reconstruction noise, in dashed and dot-dashed, true mapping of the transverse velocities will be possible.  On smaller scales, cross-correlations with large-scale structure will still be possible.  
    }
    \label{fig:ClvvNlvv}
\end{figure}


\paragraph{\textbf{Signal-to-noise ratio}}

We now estimate the signal-to-noise ratio of the reconstructed transverse velocity potential assuming a cosmology consistent with the latest results from \textit{Planck}~\citep{Aghanim:2018eyx}. We use the lensed $C_\ell^{\Theta\Theta}$, and add contributions from the kinetic Sunyaev-Zeldovich (kSZ) effect, 
which we take as a constant $3\,\mu \mathrm{K}^2$ in $\ell(\ell+1) C_\ell^{TT}/(2\pi)$~\citep{Shaw:2011sy,George:2014oba}. For the CMB temperature noise, we take ${N_\ell^{\Theta\Theta} = (\Delta_T^2/T^2) \exp[\ell(\ell+1)\theta_{\mathrm{fwhm}}^2/(8\log2)]}$. We show results for a range of CMB noise levels $\Delta_T\in[0.1,14] \ \mu$K-arcminute and beam sizes $\theta_{\mathrm{fwhm}}\in\{0.1,1.4,5.0\}$~arcminute.  
 The noise power for the moving lens potential in each redshift bin is obtained from the galaxy shot noise using the analytic approximation for the galaxy number densities ${\dd n / \dd z \propto ({z}/{z_0})^\alpha\exp[({-{z}/{z_0}})^\beta]\ \mathrm{arcmin}^{-2}}$ with $\{z_0,\alpha,\beta,n_\mathrm{tot}[\mathrm{arcmin^{-2}}]\}$ taken to be $\{0.3,2,1,40\}$ and $\{0.88,1.25,2.29,12\}$ for LSST~\citep{Abell:2009aa} and DES~\citep{2005astro.ph.10346T}, respectively.
We choose the redshift binning taking into account the photometric error expected by the these experiments, $\sigma_z=0.03(1+z)$, with each redshift bin width fixed to $4\sigma_z$, which amounts to 13 bins in the range $z\in[0,3.7]$. Finally, we assume constant galaxy bias of unity between galaxy and the matter over-density.
The moving lens potential power spectrum $C_\ell^{\psi\psi}$ is calculated with a non-linear matter power spectrum and using the Limber approximation which is valid at small scales~\cite{Limber:1953abc,Kaiser:1991qi,Kaiser:1996tp,LoVerde:2008re}. All spectra were computed numerically using modified versions of both \textsf{CAMB}~\cite{Lewis:1999bs} and  \textsf{CLASS}~\citep{2011JCAP...07..034B} with non-linear corrections implemented with \textsf{HALOFIT}~\citep{Takahashi:2012em,Mead:2015yca,Mead:2016zqy,Smith:2002dz}, and we checked that the results from the two codes agree with one another and also with the 
halo model treatment of the matter power described in~\citep{2018KSetallkSZ}.  We show the transverse velocity signal and the estimator noise in Fig.~\ref{fig:ClvvNlvv}.

The most promising route for a first detection of the moving lens effect comes from cross-correlating the large-scale transverse velocity reconstructed from the CMB with that inferred directly from a galaxy survey.  We assume that the latter method provides a precise enough measurement of the large-scale density that we can infer the large-scale transverse velocity without noise, which should be a reasonable approximation for the high number densities of galaxies expected in the surveys we are considering.
We calculate the total signal-to-noise ratio by approximating the likelihood as Gaussian 
\begin{align}
 \left(\frac{S}{N}\right)^2=&\sum_{\ell\ell'; XYWZ} C_\ell^{\vpot_X\hat{\vpot}_Y} \non
 &\times\textbf{cov}^{-1}\left(\tilde{C}^{\vpot_X\hat{\vpot}_Y}_{\ell},\tilde{C}^{\vpot_W\hat{\vpot}_Z}_{\ell'}\right) C_{\ell'}^{\vpot_W\hat{\vpot}_Z}  \, ,
\end{align}
where the indices run over redshift bins, the fields with hats refer to transverse velocities reconstructed from the CMB, those without a hat refer to the velocities reconstructed from the galaxy distribution, the tilde refers to spectra including noise, and the covariance is given by  
\begin{align}\label{eq:covdef}
 &\textbf{cov}\left(\tilde{C}^{\vpot_X\hat{\vpot}_Y}_{\ell},\tilde{C}^{\vpot_W\hat{\vpot}_Z}_{\ell'}\right) =  \frac{\delta_{\ell\ell'}}{2\ell+1} f_\mathrm{sky}^{-1} \non 
 &\qquad\times \left(\tilde{C}_{\ell}^{\vpot_X\vpot_W}\tilde{C}_{\ell}^{\hat{\vpot}_Y\hat{\vpot}_Z}+\tilde{C}_{\ell}^{\vpot_X\hat{\vpot}_Z}\tilde{C}_{\ell}^{\vpot_Y\hat{\vpot}_W}\right)  \, .
\end{align}
To assess the detectability of the moving lens effect, we take as a null hypothesis a scenario in which there is no signal in the CMB-reconstructed transverse velocity, which we also take to have noise diagonal in the redshift bins ($\tilde{C}_\ell^{\hat{X}\hat{Y}}=\delta_{\hat{X}\hat{Y}}N_\ell^{\hat{X}}$), and no signal or noise in the cross with the galaxy-derived transverse velocity ($\tilde{C}_\ell^{X\hat{Y}}=0$) when calculating the covariance matrix. 

\begin{figure}[t!]
    \includegraphics[width = \columnwidth]{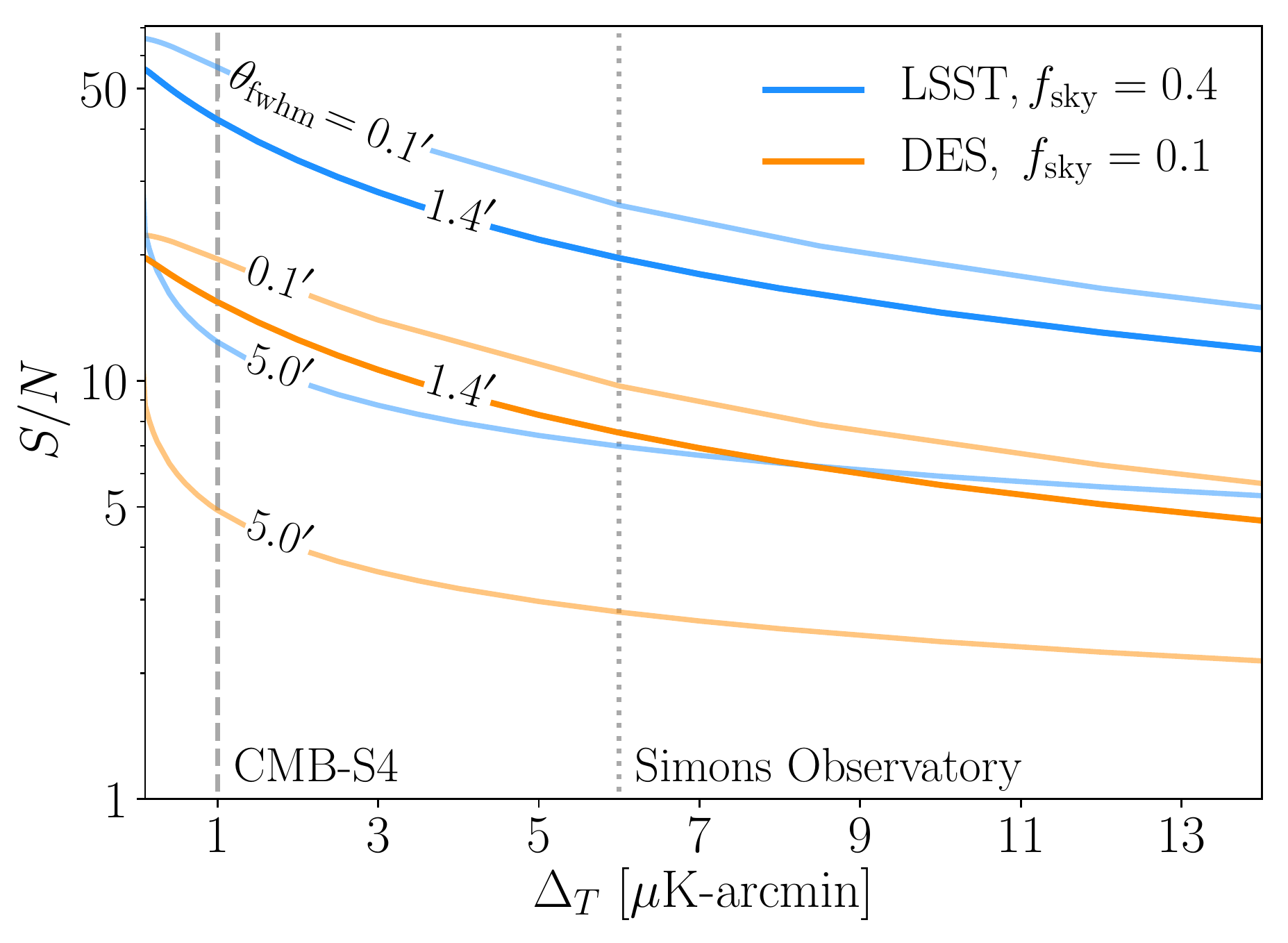}
    \caption{Signal-to-noise ratio of the transverse velocity estimator for a range of CMB noise levels and beam sizes, combined with LSST and DES.  The approximate anticipated noise levels of Simons Observatory and CMB-S4 are shown;  
    both have roughly a 1.4-arcminute beam. 
    }
    \label{fig:signal_to_noise_gold_alternative}
\end{figure}

The results for the signal-to-noise ratio with these assumptions are shown in Fig~\ref{fig:signal_to_noise_gold_alternative}.  
We find that with the method we described,
Simons Observatory combined with DES will be able to detect the moving lens effect at about 8$\sigma$, and CMB-S4 combined with LSST at about 40$\sigma$, meaning that a first detection and subsequent precision measurement of the moving lens effect should be possible in the next several years.
The signal-to-noise ratios in the results we have shown are limited in part by the contributions to the temperature spectrum that come from the kSZ effect and lensing on small scales.  Reconstructing and removing the fluctuations from the kSZ effect, which may be possible with the upcoming experiments~\citep{Deutsch:2017ybc,Smith:2018bpn}, together with applications of delensing such as in~\citep{Green:2016cjr} may improve the signal-to-noise ratio.


\paragraph{\textbf{Biases}} Ordinary lensing introduces two biases to the transverse velocity estimator. The first bias is proportional to the long-wavelength temperature gradient and takes the form 
\begin{align}\label{eq:bias1}
 \vpot^{\phi\psi}(\mathbf{L}) 
\simeq \Theta(\mathbf{L})N(\mathbf{L})\int\frac{\textnormal{d}^2\boldsymbol{\ell}}{(2\pi)^2}C^{\phi\psi}_{|\boldsymbol{\ell}-\mathbf{L}|}g(\boldsymbol{\ell},\mathbf{L})\mathbf{L}\cdot(\mathbf{L}-\boldsymbol{\ell})\, , 
\end{align}
where we have approximated the change to the temperature fluctuations due to lensing to first order in the deflection as $\Delta \Theta(\nhat)\big|_\textnormal{lens} \simeq \nabla \Theta(\nhat) \cdot \boldsymbol{\alpha}(\nhat)$.
There exists a second bias from ordinary lensing, 
\begin{align}\label{eq:bias2}
 \vpot^\mathrm{\Theta\psi}(\mathbf{L}) 
\simeq \phi(\mathbf{L})N(\mathbf{L})\int\frac{\textnormal{d}^2\boldsymbol{\ell}}{(2\pi)^2}C^{\Theta\psi}_{|\boldsymbol{\ell}-\mathbf{L}|}g(\boldsymbol{\ell},\mathbf{L})\mathbf{L}\cdot(\mathbf{L}-\boldsymbol{\ell})\, ,
\end{align}
which can be understood as the large-scale gravitational potential fluctuations distorting small-scale ISW or Rees-Sciama temperature fluctuations.

The kSZ effect generates CMB temperature fluctuations of the form {$\Delta \Theta(\nhat)\big|_{\textnormal{kSZ}}=-\int\textnormal{d}\chi\; {v}_\textnormal{d}(\chi\nhat)\, d\tau/d\chi(\chi\nhat)$
where $d\tau/d\chi(\chi\nhat)=\sigma_T a  n_e(\chi\nhat)$,}  
$\sigma_T$ is the Thomson cross section, $a$ is the scale factor, $n_e$ is the free electron number density, and $v_\textnormal{d}$ is the remote CMB dipole projected along the line of sight, given by $v_\textnormal{d}=3\int\textnormal{d}^2\hat{\mathbf{n}}\;\Theta_1(\hat{\mathbf{n}}_e,\hat{\mathbf{n}})(\hat{\mathbf{n}}_e\cdot\hat{\mathbf{n}})/(4\pi)$. 
We approximate the dipole seen by distant electrons as dominated by the Doppler effect $\Theta_1\simeq\mathbf{v}_e\cdot\hat{\mathbf{n}}$,  where $\mathbf{v}_e$ is the electron velocity.  The contribution from the kSZ effect to our transverse velocity estimator is then 
\begin{align}\label{eq:kSZ_cont_est}
 \vpot^\mathrm{kSZ}(\mathbf{L}) 
\simeq -v_\textnormal{d}(\mathbf{L})N(\mathbf{L}) \!\int\frac{\textnormal{d}^2\boldsymbol{\ell}}{(2\pi)^2}C^{\delta\tau\psi}_{|\boldsymbol{\ell}-\mathbf{L}|}g(\boldsymbol{\ell},\mathbf{L})\ ,
\end{align}
where 
$C^{\delta\tau\psi}_{\ell}$ is the cross-correlation between $\psi$ and $d\tau/d\chi$.

\begin{figure}[t!]
    \includegraphics[width = \columnwidth]{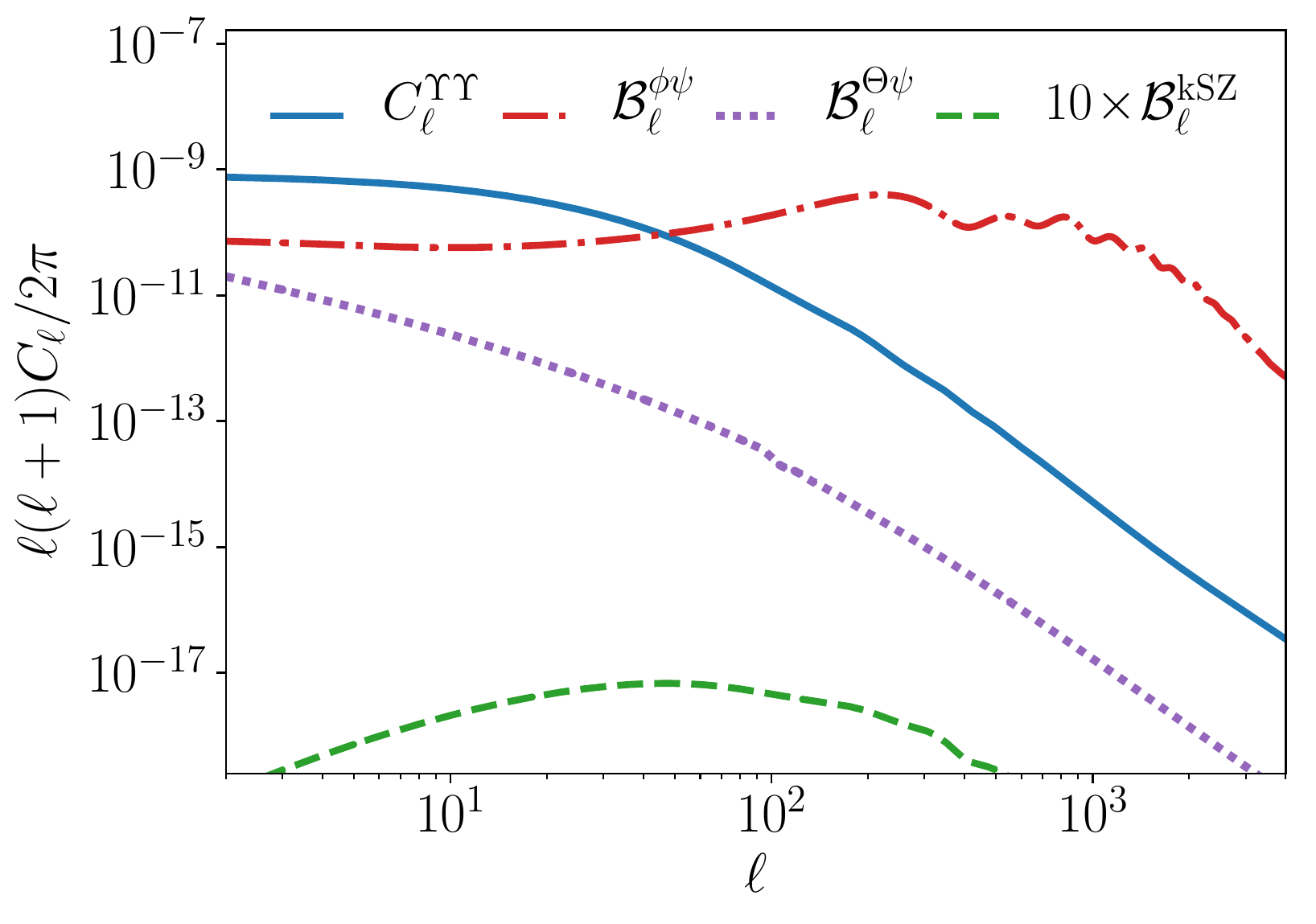}
    \caption{Comparison of the transverse velocity power spectrum with ordinary lensing and kSZ biases for the redshift bin $z\in[1.00,1.25]$ for a CMB experiment with $\Delta_T=1\,\mu \mathrm{K}$-arcmin and a 1.4-arcmin beam combined with LSST. The dominant contribution to the signal-to-noise ratio comes from large scales $\ell \lesssim  50$, where the biases are smaller than the transverse velocity signal.  Furthermore these biases can be mitigated using the methods described in the main text.
    } 
    \label{fig:signal-to-noise_bias_comparison}
\end{figure}

We now assess how large these biases would be if one were to naively apply the estimator shown in Eq.~\eqref{eq:noise_naive} to the data.  We define the spectra of the biases as ${\langle \vpot^B(\vl{})\vpot^B(\vlp{})\rangle=(2\pi)^2\mathcal{B}^B_\ell\delta^{(2)}(\vl{}+\vlp{})}$ where $B\in\{\phi\psi,\Theta\psi,\mathrm{kSZ}\}$ and plot the results in Fig.~\ref{fig:signal-to-noise_bias_comparison} for the redshift bin $z\in[1.00,1.25]$. One can see that the $\phi\psi$-bias introduced in Eq.~\eqref{eq:bias1} traces the structure of the primary CMB temperature, due to the fact that our transverse velocity estimator is very similar to an estimator designed to reconstruct the large-scale primary temperature fluctuations from observation of small-scale temperature and lenses~\citep{Foreman:2018lci}. This bias is the largest of those we have considered, and it is smaller than the signal on large scales $\ell\lesssim50$ which make the dominant contribution to the signal-to-noise ratio. Our knowledge of the large-scale CMB temperature allows us to cleanly remove the effects of the $\psi\phi$-bias by subtracting a best-fit multiple of the observed large scale temperature fluctuations from the reconstructed $\Upsilon$ map.  This bias could also be reduced by delensing the temperature map~\citep{Hirata:2002jy,Green:2016cjr,Millea:2017fyd} before estimating the transverse velocity potential, or by suppressing its contribution to the estimator by bias-hardening~\citep{namikawa2013bias}.

The $\Theta\psi$-bias introduced in Eq.~\eqref{eq:bias2} is most important on large scales, though it is about two orders of magnitude smaller than the transverse velocity signal on most scales and redshifts. Our estimate of this bias included only the linear contributions to the ISW effect, but the non-linear Rees-Sciama effect may increase $C_\ell^{\Theta\psi}$ on small scales, thereby boosting the bias compared to what we have calculated here. The $\Theta\psi$-bias can also be mitigated by subtracting from the reconstructed $\Upsilon$ map the best-fit multiple of the gravitational lensing field $\phi$ which will be measured at high significance with the CMB experiments we are considering. The kSZ bias is sub-dominant on all scales of interest, though it too may be possible to reconstruct and remove with the experiments being discussed here~\citep{Deutsch:2017ybc,Smith:2018bpn}.


\paragraph{\textbf{Discussion}}

It has long been known that gravitational potentials moving transverse to our line of sight generate temperature fluctuations in the CMB~\cite{1983Natur.302..315B}. Individual objects induce small fluctuations in the temperature which are easily confused with other effects making detection challenging~\citep{Tuluie:1995ut,Aghanim:1998ux,Cooray:2002ee}.  
By statistically combining the signal from many objects with a common bulk motion, the method we described greatly increases the prospects for reconstructing transverse velocities on large scales.
We demonstrated that upcoming CMB experiments like Simons Observatory and CMB-S4 combined with galaxy surveys such as DES and LSST have the statistical power to make a detection of the moving lens effect at high significance.
We also computed the leading biases and discussed how they can be mitigated.

Using the CMB to reconstruct the large-scale transverse velocity field allows for the use of small-scale CMB measurements to probe long-wavelength cosmological fluctuations at lower redshift, much like with CMB lensing~\citep{Hu:2001kj}, the kSZ effect~\citep{Zhang10d,Deutsch:2017ybc,Smith:2018bpn}, and the polarized SZ effect~\citep{Kamionkowski:1997na,Bunn:2006mp,Deutsch:2017cja,Meyers:2017rtf}. 
Since the observation of large-scale modes is typically challenging, and the number of independent modes on large scales is inherently limited, it is generally useful to expand the list of methods to access large scales observationally.  As a specific application, one could imagine using the large-scale velocity modes reconstructed with the moving lens effect to cancel cosmic variance~\citep{Seljak:2008xr} for the purpose of constraining local non-Gaussianity (which induces a scale-dependent bias on large scales~\citep{Dalal:2007cu}), in a way similar to what has been explored for CMB lensing~\citep{Schmittfull:2017ffw} and the kSZ effect~\citep{Munchmeyer:2018eey}. Furthermore, because the moving lens effect is a result of purely gravitational effects, it can be used to measure quantities which cannot be accessed directly with the kSZ effect alone, such as the absolute growth rate, which is useful for studying dark energy~\citep{Linder:2005in}, modified gravity~\citep{Linder:2007hg}, and the effects of neutrino mass~\citep{Bond:1983hb}.  Combined with other probes, observations of the moving lens effect can also help reduce degeneracies due to astrophysical uncertainties such as the optical depth degeneracy of the kSZ effect~\citep{Smith:2018bpn}.

The major leaps forward in the precision of near-future CMB and galaxy surveys will open many new cosmological opportunities.  We have described a method which will allow for the first detection of the moving lens effect with forthcoming data, and will provide a novel probe of large-scale transverse velocities with a host of cosmological applications.


\paragraph{\textbf{Acknowledgments}}

We would like to thank Carlo Contaldi, Alan Heavens, Emmanuel Schaan, and David Spergel for helpful discussions. Research at Perimeter Institute is supported by the Government of Canada through Industry Canada and by the Province of Ontario through the Ministry of Research \& Innovation. MCJ is supported by the National Science and Engineering Research Council through a Discovery grant. SCH is funded by the Imperial College President's Scholarship.   

\bibliography{moving_lens}

\begin{thebibliography}{61}%
\makeatletter
\providecommand \@ifxundefined [1]{%
 \@ifx{#1\undefined}
}%
\providecommand \@ifnum [1]{%
 \ifnum #1\expandafter \@firstoftwo
 \else \expandafter \@secondoftwo
 \fi
}%
\providecommand \@ifx [1]{%
 \ifx #1\expandafter \@firstoftwo
 \else \expandafter \@secondoftwo
 \fi
}%
\providecommand \natexlab [1]{#1}%
\providecommand \enquote  [1]{``#1''}%
\providecommand \bibnamefont  [1]{#1}%
\providecommand \bibfnamefont [1]{#1}%
\providecommand \citenamefont [1]{#1}%
\providecommand \href@noop [0]{\@secondoftwo}%
\providecommand \href [0]{\begingroup \@sanitize@url \@href}%
\providecommand \@href[1]{\@@startlink{#1}\@@href}%
\providecommand \@@href[1]{\endgroup#1\@@endlink}%
\providecommand \@sanitize@url [0]{\catcode `\\12\catcode `\$12\catcode
  `\&12\catcode `\#12\catcode `\^12\catcode `\_12\catcode `\%12\relax}%
\providecommand \@@startlink[1]{}%
\providecommand \@@endlink[0]{}%
\providecommand \url  [0]{\begingroup\@sanitize@url \@url }%
\providecommand \@url [1]{\endgroup\@href {#1}{\urlprefix }}%
\providecommand \urlprefix  [0]{URL }%
\providecommand \Eprint [0]{\href }%
\providecommand \doibase [0]{http://dx.doi.org/}%
\providecommand \selectlanguage [0]{\@gobble}%
\providecommand \bibinfo  [0]{\@secondoftwo}%
\providecommand \bibfield  [0]{\@secondoftwo}%
\providecommand \translation [1]{[#1]}%
\providecommand \BibitemOpen [0]{}%
\providecommand \bibitemStop [0]{}%
\providecommand \bibitemNoStop [0]{.\EOS\space}%
\providecommand \EOS [0]{\spacefactor3000\relax}%
\providecommand \BibitemShut  [1]{\csname bibitem#1\endcsname}%
\let\auto@bib@innerbib\@empty
\bibitem [{\citenamefont {Aguirre}\ \emph {et~al.}(2018)\citenamefont {Aguirre}
  \emph {et~al.}}]{Ade:2018sbj}%
  \BibitemOpen
  \bibfield  {author} {\bibinfo {author} {\bibfnamefont {J.}~\bibnamefont
  {Aguirre}} \emph {et~al.} (\bibinfo {collaboration} {Simons Observatory}),\
  }\href@noop {} {\  (\bibinfo {year} {2018})},\ \Eprint
  {http://arxiv.org/abs/1808.07445} {arXiv:1808.07445 [astro-ph.CO]}
  \BibitemShut {NoStop}%
\bibitem [{\citenamefont {Abazajian}\ \emph {et~al.}(2016)\citenamefont
  {Abazajian} \emph {et~al.}}]{Abazajian:2016yjj}%
  \BibitemOpen
  \bibfield  {author} {\bibinfo {author} {\bibfnamefont {K.~N.}\ \bibnamefont
  {Abazajian}} \emph {et~al.} (\bibinfo {collaboration} {CMB-S4}),\ }\href@noop
  {} {\  (\bibinfo {year} {2016})},\ \Eprint {http://arxiv.org/abs/1610.02743}
  {arXiv:1610.02743 [astro-ph.CO]} \BibitemShut {NoStop}%
\bibitem [{\citenamefont {{The Dark Energy Survey
  Collaboration}}(2005)}]{2005astro.ph.10346T}%
  \BibitemOpen
  \bibfield  {author} {\bibinfo {author} {\bibnamefont {{The Dark Energy Survey
  Collaboration}}},\ }\href@noop {} {\bibfield  {journal} {\bibinfo  {journal}
  {ArXiv Astrophysics e-prints}\ } (\bibinfo {year} {2005})},\ \Eprint
  {http://arxiv.org/abs/astro-ph/0510346} {astro-ph/0510346} \BibitemShut
  {NoStop}%
\bibitem [{\citenamefont {Abell}\ \emph {et~al.}(2009)\citenamefont {Abell}
  \emph {et~al.}}]{Abell:2009aa}%
  \BibitemOpen
  \bibfield  {author} {\bibinfo {author} {\bibfnamefont {P.~A.}\ \bibnamefont
  {Abell}} \emph {et~al.} (\bibinfo {collaboration} {LSST Science, LSST
  Project}),\ }\href@noop {} {\  (\bibinfo {year} {2009})},\ \Eprint
  {http://arxiv.org/abs/0912.0201} {arXiv:0912.0201 [astro-ph.IM]} \BibitemShut
  {NoStop}%
\bibitem [{\citenamefont {Lewis}\ and\ \citenamefont
  {Challinor}(2006)}]{Lewis:2006fu}%
  \BibitemOpen
  \bibfield  {author} {\bibinfo {author} {\bibfnamefont {A.}~\bibnamefont
  {Lewis}}\ and\ \bibinfo {author} {\bibfnamefont {A.}~\bibnamefont
  {Challinor}},\ }\href {\doibase 10.1016/j.physrep.2006.03.002} {\bibfield
  {journal} {\bibinfo  {journal} {Phys. Rept.}\ }\textbf {\bibinfo {volume}
  {429}},\ \bibinfo {pages} {1} (\bibinfo {year} {2006})},\ \Eprint
  {http://arxiv.org/abs/astro-ph/0601594} {arXiv:astro-ph/0601594 [astro-ph]}
  \BibitemShut {NoStop}%
\bibitem [{\citenamefont {{Sachs}}\ and\ \citenamefont
  {{Wolfe}}(1967)}]{1967ApJ...147...73S}%
  \BibitemOpen
  \bibfield  {author} {\bibinfo {author} {\bibfnamefont {R.~K.}\ \bibnamefont
  {{Sachs}}}\ and\ \bibinfo {author} {\bibfnamefont {A.~M.}\ \bibnamefont
  {{Wolfe}}},\ }\href {\doibase 10.1086/148982} {\bibfield  {journal} {\bibinfo
   {journal} {\apj}\ }\textbf {\bibinfo {volume} {147}},\ \bibinfo {pages} {73}
  (\bibinfo {year} {1967})}\BibitemShut {NoStop}%
\bibitem [{\citenamefont {{Rees}}\ and\ \citenamefont
  {{Sciama}}(1968)}]{1968Natur.217..511R}%
  \BibitemOpen
  \bibfield  {author} {\bibinfo {author} {\bibfnamefont {M.~J.}\ \bibnamefont
  {{Rees}}}\ and\ \bibinfo {author} {\bibfnamefont {D.~W.}\ \bibnamefont
  {{Sciama}}},\ }\href {\doibase 10.1038/217511a0} {\bibfield  {journal}
  {\bibinfo  {journal} {\nat}\ }\textbf {\bibinfo {volume} {217}},\ \bibinfo
  {pages} {511} (\bibinfo {year} {1968})}\BibitemShut {NoStop}%
\bibitem [{\citenamefont {Zeldovich}\ and\ \citenamefont
  {Sunyaev}(1969)}]{Zeldovich:1969ff}%
  \BibitemOpen
  \bibfield  {author} {\bibinfo {author} {\bibfnamefont {{\relax Ya}.~B.}\
  \bibnamefont {Zeldovich}}\ and\ \bibinfo {author} {\bibfnamefont {R.~A.}\
  \bibnamefont {Sunyaev}},\ }\href {\doibase 10.1007/BF00661821} {\bibfield
  {journal} {\bibinfo  {journal} {Astrophys. Space Sci.}\ }\textbf {\bibinfo
  {volume} {4}},\ \bibinfo {pages} {301} (\bibinfo {year} {1969})}\BibitemShut
  {NoStop}%
\bibitem [{\citenamefont {Sunyaev}\ and\ \citenamefont
  {Zeldovich}(1970)}]{Sunyaev:1970er}%
  \BibitemOpen
  \bibfield  {author} {\bibinfo {author} {\bibfnamefont {R.~A.}\ \bibnamefont
  {Sunyaev}}\ and\ \bibinfo {author} {\bibfnamefont {{\relax Ya}.~B.}\
  \bibnamefont {Zeldovich}},\ }\href@noop {} {\bibfield  {journal} {\bibinfo
  {journal} {Astrophys. Space Sci.}\ }\textbf {\bibinfo {volume} {7}},\
  \bibinfo {pages} {20} (\bibinfo {year} {1970})}\BibitemShut {NoStop}%
\bibitem [{\citenamefont {Sunyaev}\ and\ \citenamefont
  {Zeldovich}(1972)}]{Sunyaev:1972eq}%
  \BibitemOpen
  \bibfield  {author} {\bibinfo {author} {\bibfnamefont {R.~A.}\ \bibnamefont
  {Sunyaev}}\ and\ \bibinfo {author} {\bibfnamefont {{\relax Ya}.~B.}\
  \bibnamefont {Zeldovich}},\ }\href@noop {} {\bibfield  {journal} {\bibinfo
  {journal} {Comments Astrophys. Space Phys.}\ }\textbf {\bibinfo {volume}
  {4}},\ \bibinfo {pages} {173} (\bibinfo {year} {1972})}\BibitemShut {NoStop}%
\bibitem [{\citenamefont {Sunyaev}\ and\ \citenamefont
  {Zeldovich}(1980)}]{Sunyaev:1980vz}%
  \BibitemOpen
  \bibfield  {author} {\bibinfo {author} {\bibfnamefont {R.~A.}\ \bibnamefont
  {Sunyaev}}\ and\ \bibinfo {author} {\bibfnamefont {{\relax Ya}.~B.}\
  \bibnamefont {Zeldovich}},\ }\href {\doibase
  10.1146/annurev.aa.18.090180.002541} {\bibfield  {journal} {\bibinfo
  {journal} {Ann. Rev. Astron. Astrophys.}\ }\textbf {\bibinfo {volume} {18}},\
  \bibinfo {pages} {537} (\bibinfo {year} {1980})}\BibitemShut {NoStop}%
\bibitem [{\citenamefont {Sazonov}\ and\ \citenamefont
  {Sunyaev}(1999)}]{Sazonov:1999zp}%
  \BibitemOpen
  \bibfield  {author} {\bibinfo {author} {\bibfnamefont {S.~Y.}\ \bibnamefont
  {Sazonov}}\ and\ \bibinfo {author} {\bibfnamefont {R.~A.}\ \bibnamefont
  {Sunyaev}},\ }\href {\doibase 10.1046/j.1365-8711.1999.02981.x} {\bibfield
  {journal} {\bibinfo  {journal} {Mon. Not. Roy. Astron. Soc.}\ }\textbf
  {\bibinfo {volume} {310}},\ \bibinfo {pages} {765} (\bibinfo {year}
  {1999})},\ \Eprint {http://arxiv.org/abs/astro-ph/9903287}
  {arXiv:astro-ph/9903287 [astro-ph]} \BibitemShut {NoStop}%
\bibitem [{\citenamefont {{Birkinshaw}}\ and\ \citenamefont
  {{Gull}}(1983)}]{1983Natur.302..315B}%
  \BibitemOpen
  \bibfield  {author} {\bibinfo {author} {\bibfnamefont {M.}~\bibnamefont
  {{Birkinshaw}}}\ and\ \bibinfo {author} {\bibfnamefont {S.~F.}\ \bibnamefont
  {{Gull}}},\ }\href {\doibase 10.1038/302315a0} {\bibfield  {journal}
  {\bibinfo  {journal} {\nat}\ }\textbf {\bibinfo {volume} {302}},\ \bibinfo
  {pages} {315} (\bibinfo {year} {1983})}\BibitemShut {NoStop}%
\bibitem [{\citenamefont {Tuluie}\ and\ \citenamefont
  {Laguna}(1995)}]{Tuluie:1995ut}%
  \BibitemOpen
  \bibfield  {author} {\bibinfo {author} {\bibfnamefont {R.}~\bibnamefont
  {Tuluie}}\ and\ \bibinfo {author} {\bibfnamefont {P.}~\bibnamefont
  {Laguna}},\ }\href {\doibase 10.1086/187893} {\bibfield  {journal} {\bibinfo
  {journal} {Astrophys. J.}\ }\textbf {\bibinfo {volume} {445}},\ \bibinfo
  {pages} {L73} (\bibinfo {year} {1995})},\ \Eprint
  {http://arxiv.org/abs/astro-ph/9501059} {arXiv:astro-ph/9501059 [astro-ph]}
  \BibitemShut {NoStop}%
\bibitem [{\citenamefont {Aghanim}\ \emph {et~al.}(1998)\citenamefont
  {Aghanim}, \citenamefont {Prunet}, \citenamefont {Forni},\ and\ \citenamefont
  {Bouchet}}]{Aghanim:1998ux}%
  \BibitemOpen
  \bibfield  {author} {\bibinfo {author} {\bibfnamefont {N.}~\bibnamefont
  {Aghanim}}, \bibinfo {author} {\bibfnamefont {S.}~\bibnamefont {Prunet}},
  \bibinfo {author} {\bibfnamefont {O.}~\bibnamefont {Forni}}, \ and\ \bibinfo
  {author} {\bibfnamefont {F.~R.}\ \bibnamefont {Bouchet}},\ }\href@noop {}
  {\bibfield  {journal} {\bibinfo  {journal} {Submitted to: Astron.
  Astrophys.}\ } (\bibinfo {year} {1998})},\ \bibinfo {note} {[Astron.
  Astrophys.334,409(1998)]},\ \Eprint {http://arxiv.org/abs/astro-ph/9803040}
  {arXiv:astro-ph/9803040 [astro-ph]} \BibitemShut {NoStop}%
\bibitem [{\citenamefont {Cooray}(2002)}]{Cooray:2002ee}%
  \BibitemOpen
  \bibfield  {author} {\bibinfo {author} {\bibfnamefont {A.}~\bibnamefont
  {Cooray}},\ }\href {\doibase 10.1103/PhysRevD.65.083518} {\bibfield
  {journal} {\bibinfo  {journal} {Phys. Rev.}\ }\textbf {\bibinfo {volume}
  {D65}},\ \bibinfo {pages} {083518} (\bibinfo {year} {2002})},\ \Eprint
  {http://arxiv.org/abs/astro-ph/0109162} {arXiv:astro-ph/0109162 [astro-ph]}
  \BibitemShut {NoStop}%
\bibitem [{\citenamefont {{Aghanim}}\ \emph {et~al.}(1998)\citenamefont
  {{Aghanim}}, \citenamefont {{Prunet}}, \citenamefont {{Forni}},\ and\
  \citenamefont {{Bouchet}}}]{1998A&A...334..409A}%
  \BibitemOpen
  \bibfield  {author} {\bibinfo {author} {\bibfnamefont {N.}~\bibnamefont
  {{Aghanim}}}, \bibinfo {author} {\bibfnamefont {S.}~\bibnamefont {{Prunet}}},
  \bibinfo {author} {\bibfnamefont {O.}~\bibnamefont {{Forni}}}, \ and\
  \bibinfo {author} {\bibfnamefont {F.~R.}\ \bibnamefont {{Bouchet}}},\
  }\href@noop {} {\bibfield  {journal} {\bibinfo  {journal} {\aap}\ }\textbf
  {\bibinfo {volume} {334}},\ \bibinfo {pages} {409} (\bibinfo {year}
  {1998})},\ \Eprint {http://arxiv.org/abs/astro-ph/9803040} {astro-ph/9803040}
  \BibitemShut {NoStop}%
\bibitem [{\citenamefont {{Gurvits}}\ and\ \citenamefont
  {{Mitrofanov}}(1986)}]{Gurvitz:1986ab}%
  \BibitemOpen
  \bibfield  {author} {\bibinfo {author} {\bibfnamefont {L.~I.}\ \bibnamefont
  {{Gurvits}}}\ and\ \bibinfo {author} {\bibfnamefont {I.~G.}\ \bibnamefont
  {{Mitrofanov}}},\ }\href {\doibase 10.1038/324349a0} {\bibfield  {journal}
  {\bibinfo  {journal} {\nat}\ }\textbf {\bibinfo {volume} {324}},\ \bibinfo
  {pages} {349} (\bibinfo {year} {1986})}\BibitemShut {NoStop}%
\bibitem [{\citenamefont {{Cooray}}(2002)}]{2002PhRvD..65h3518C}%
  \BibitemOpen
  \bibfield  {author} {\bibinfo {author} {\bibfnamefont {A.}~\bibnamefont
  {{Cooray}}},\ }\href {\doibase 10.1103/PhysRevD.65.083518} {\bibfield
  {journal} {\bibinfo  {journal} {\prd}\ }\textbf {\bibinfo {volume} {65}},\
  \bibinfo {eid} {083518} (\bibinfo {year} {2002})},\ \Eprint
  {http://arxiv.org/abs/astro-ph/0109162} {astro-ph/0109162} \BibitemShut
  {NoStop}%
\bibitem [{\citenamefont {Challinor}\ and\ \citenamefont {van
  Leeuwen}(2002)}]{Challinor:2002zh}%
  \BibitemOpen
  \bibfield  {author} {\bibinfo {author} {\bibfnamefont {A.}~\bibnamefont
  {Challinor}}\ and\ \bibinfo {author} {\bibfnamefont {F.}~\bibnamefont {van
  Leeuwen}},\ }\href {\doibase 10.1103/PhysRevD.65.103001} {\bibfield
  {journal} {\bibinfo  {journal} {Phys. Rev.}\ }\textbf {\bibinfo {volume}
  {D65}},\ \bibinfo {pages} {103001} (\bibinfo {year} {2002})},\ \Eprint
  {http://arxiv.org/abs/astro-ph/0112457} {arXiv:astro-ph/0112457 [astro-ph]}
  \BibitemShut {NoStop}%
\bibitem [{\citenamefont {Aghanim}\ \emph {et~al.}(2014)\citenamefont {Aghanim}
  \emph {et~al.}}]{Aghanim:2013suk}%
  \BibitemOpen
  \bibfield  {author} {\bibinfo {author} {\bibfnamefont {N.}~\bibnamefont
  {Aghanim}} \emph {et~al.} (\bibinfo {collaboration} {Planck}),\ }\href
  {\doibase 10.1051/0004-6361/201321556} {\bibfield  {journal} {\bibinfo
  {journal} {Astron. Astrophys.}\ }\textbf {\bibinfo {volume} {571}},\ \bibinfo
  {pages} {A27} (\bibinfo {year} {2014})},\ \Eprint
  {http://arxiv.org/abs/1303.5087} {arXiv:1303.5087 [astro-ph.CO]} \BibitemShut
  {NoStop}%
\bibitem [{\citenamefont {Aghanim}\ \emph
  {et~al.}(2018{\natexlab{a}})\citenamefont {Aghanim} \emph
  {et~al.}}]{Aghanim:2018fcm}%
  \BibitemOpen
  \bibfield  {author} {\bibinfo {author} {\bibfnamefont {N.}~\bibnamefont
  {Aghanim}} \emph {et~al.} (\bibinfo {collaboration} {Planck}),\ }\href@noop
  {} {\  (\bibinfo {year} {2018}{\natexlab{a}})},\ \Eprint
  {http://arxiv.org/abs/1807.06207} {arXiv:1807.06207 [astro-ph.CO]}
  \BibitemShut {NoStop}%
\bibitem [{\citenamefont {Notari}\ and\ \citenamefont
  {Quartin}(2012)}]{Notari:2011sb}%
  \BibitemOpen
  \bibfield  {author} {\bibinfo {author} {\bibfnamefont {A.}~\bibnamefont
  {Notari}}\ and\ \bibinfo {author} {\bibfnamefont {M.}~\bibnamefont
  {Quartin}},\ }\href {\doibase 10.1088/1475-7516/2012/02/026} {\bibfield
  {journal} {\bibinfo  {journal} {JCAP}\ }\textbf {\bibinfo {volume} {1202}},\
  \bibinfo {pages} {026} (\bibinfo {year} {2012})},\ \Eprint
  {http://arxiv.org/abs/1112.1400} {arXiv:1112.1400 [astro-ph.CO]} \BibitemShut
  {NoStop}%
\bibitem [{\citenamefont {Lewis}\ \emph {et~al.}(2017)\citenamefont {Lewis},
  \citenamefont {Hall},\ and\ \citenamefont {Challinor}}]{Lewis:2017ans}%
  \BibitemOpen
  \bibfield  {author} {\bibinfo {author} {\bibfnamefont {A.}~\bibnamefont
  {Lewis}}, \bibinfo {author} {\bibfnamefont {A.}~\bibnamefont {Hall}}, \ and\
  \bibinfo {author} {\bibfnamefont {A.}~\bibnamefont {Challinor}},\ }\href
  {\doibase 10.1088/1475-7516/2017/08/023} {\bibfield  {journal} {\bibinfo
  {journal} {JCAP}\ }\textbf {\bibinfo {volume} {1708}},\ \bibinfo {pages}
  {023} (\bibinfo {year} {2017})},\ \Eprint {http://arxiv.org/abs/1706.02673}
  {arXiv:1706.02673 [astro-ph.CO]} \BibitemShut {NoStop}%
\bibitem [{\citenamefont {Cooray}\ and\ \citenamefont
  {Seto}(2005)}]{Cooray:2005my}%
  \BibitemOpen
  \bibfield  {author} {\bibinfo {author} {\bibfnamefont {A.}~\bibnamefont
  {Cooray}}\ and\ \bibinfo {author} {\bibfnamefont {N.}~\bibnamefont {Seto}},\
  }\href {\doibase 10.1088/1475-7516/2005/12/004} {\bibfield  {journal}
  {\bibinfo  {journal} {JCAP}\ }\textbf {\bibinfo {volume} {0512}},\ \bibinfo
  {pages} {004} (\bibinfo {year} {2005})},\ \Eprint
  {http://arxiv.org/abs/astro-ph/0510137} {arXiv:astro-ph/0510137 [astro-ph]}
  \BibitemShut {NoStop}%
\bibitem [{\citenamefont {Meerburg}\ \emph {et~al.}(2017)\citenamefont
  {Meerburg}, \citenamefont {Meyers},\ and\ \citenamefont {van
  Engelen}}]{Meerburg:2017xga}%
  \BibitemOpen
  \bibfield  {author} {\bibinfo {author} {\bibfnamefont {P.~D.}\ \bibnamefont
  {Meerburg}}, \bibinfo {author} {\bibfnamefont {J.}~\bibnamefont {Meyers}}, \
  and\ \bibinfo {author} {\bibfnamefont {A.}~\bibnamefont {van Engelen}},\
  }\href {\doibase 10.1103/PhysRevD.96.083519} {\bibfield  {journal} {\bibinfo
  {journal} {Phys. Rev.}\ }\textbf {\bibinfo {volume} {D96}},\ \bibinfo {pages}
  {083519} (\bibinfo {year} {2017})},\ \Eprint
  {http://arxiv.org/abs/1704.00718} {arXiv:1704.00718 [astro-ph.CO]}
  \BibitemShut {NoStop}%
\bibitem [{\citenamefont {Hu}\ and\ \citenamefont {Okamoto}(2002)}]{Hu:2001kj}%
  \BibitemOpen
  \bibfield  {author} {\bibinfo {author} {\bibfnamefont {W.}~\bibnamefont
  {Hu}}\ and\ \bibinfo {author} {\bibfnamefont {T.}~\bibnamefont {Okamoto}},\
  }\href {\doibase 10.1086/341110} {\bibfield  {journal} {\bibinfo  {journal}
  {Astrophys. J.}\ }\textbf {\bibinfo {volume} {574}},\ \bibinfo {pages} {566}
  (\bibinfo {year} {2002})},\ \Eprint {http://arxiv.org/abs/astro-ph/0111606}
  {arXiv:astro-ph/0111606 [astro-ph]} \BibitemShut {NoStop}%
\bibitem [{\citenamefont {Okamoto}\ and\ \citenamefont
  {Hu}(2003)}]{Okamoto:2003zw}%
  \BibitemOpen
  \bibfield  {author} {\bibinfo {author} {\bibfnamefont {T.}~\bibnamefont
  {Okamoto}}\ and\ \bibinfo {author} {\bibfnamefont {W.}~\bibnamefont {Hu}},\
  }\href {\doibase 10.1103/PhysRevD.67.083002} {\bibfield  {journal} {\bibinfo
  {journal} {Phys. Rev.}\ }\textbf {\bibinfo {volume} {D67}},\ \bibinfo {pages}
  {083002} (\bibinfo {year} {2003})},\ \Eprint
  {http://arxiv.org/abs/astro-ph/0301031} {arXiv:astro-ph/0301031 [astro-ph]}
  \BibitemShut {NoStop}%
\bibitem [{\citenamefont {Aghanim}\ \emph
  {et~al.}(2018{\natexlab{b}})\citenamefont {Aghanim} \emph
  {et~al.}}]{Aghanim:2018eyx}%
  \BibitemOpen
  \bibfield  {author} {\bibinfo {author} {\bibfnamefont {N.}~\bibnamefont
  {Aghanim}} \emph {et~al.} (\bibinfo {collaboration} {Planck}),\ }\href@noop
  {} {\  (\bibinfo {year} {2018}{\natexlab{b}})},\ \Eprint
  {http://arxiv.org/abs/1807.06209} {arXiv:1807.06209 [astro-ph.CO]}
  \BibitemShut {NoStop}%
\bibitem [{\citenamefont {Shaw}\ \emph {et~al.}(2012)\citenamefont {Shaw},
  \citenamefont {Rudd},\ and\ \citenamefont {Nagai}}]{Shaw:2011sy}%
  \BibitemOpen
  \bibfield  {author} {\bibinfo {author} {\bibfnamefont {L.~D.}\ \bibnamefont
  {Shaw}}, \bibinfo {author} {\bibfnamefont {D.~H.}\ \bibnamefont {Rudd}}, \
  and\ \bibinfo {author} {\bibfnamefont {D.}~\bibnamefont {Nagai}},\ }\href
  {\doibase 10.1088/0004-637X/756/1/15} {\bibfield  {journal} {\bibinfo
  {journal} {Astrophys. J.}\ }\textbf {\bibinfo {volume} {756}},\ \bibinfo
  {pages} {15} (\bibinfo {year} {2012})},\ \Eprint
  {http://arxiv.org/abs/1109.0553} {arXiv:1109.0553 [astro-ph.CO]} \BibitemShut
  {NoStop}%
\bibitem [{\citenamefont {George}\ \emph {et~al.}(2015)\citenamefont {George}
  \emph {et~al.}}]{George:2014oba}%
  \BibitemOpen
  \bibfield  {author} {\bibinfo {author} {\bibfnamefont {E.~M.}\ \bibnamefont
  {George}} \emph {et~al.},\ }\href {\doibase 10.1088/0004-637X/799/2/177}
  {\bibfield  {journal} {\bibinfo  {journal} {Astrophys. J.}\ }\textbf
  {\bibinfo {volume} {799}},\ \bibinfo {pages} {177} (\bibinfo {year}
  {2015})},\ \Eprint {http://arxiv.org/abs/1408.3161} {arXiv:1408.3161
  [astro-ph.CO]} \BibitemShut {NoStop}%
\bibitem [{\citenamefont {{Limber}}(1953)}]{Limber:1953abc}%
  \BibitemOpen
  \bibfield  {author} {\bibinfo {author} {\bibfnamefont {D.~N.}\ \bibnamefont
  {{Limber}}},\ }\href {\doibase 10.1086/145672} {\bibfield  {journal}
  {\bibinfo  {journal} {\apj}\ }\textbf {\bibinfo {volume} {117}},\ \bibinfo
  {pages} {134} (\bibinfo {year} {1953})}\BibitemShut {NoStop}%
\bibitem [{\citenamefont {Kaiser}(1992)}]{Kaiser:1991qi}%
  \BibitemOpen
  \bibfield  {author} {\bibinfo {author} {\bibfnamefont {N.}~\bibnamefont
  {Kaiser}},\ }\href {\doibase 10.1086/171151} {\bibfield  {journal} {\bibinfo
  {journal} {Astrophys. J.}\ }\textbf {\bibinfo {volume} {388}},\ \bibinfo
  {pages} {272} (\bibinfo {year} {1992})}\BibitemShut {NoStop}%
\bibitem [{\citenamefont {Kaiser}(1998)}]{Kaiser:1996tp}%
  \BibitemOpen
  \bibfield  {author} {\bibinfo {author} {\bibfnamefont {N.}~\bibnamefont
  {Kaiser}},\ }\href {\doibase 10.1086/305515} {\bibfield  {journal} {\bibinfo
  {journal} {Astrophys. J.}\ }\textbf {\bibinfo {volume} {498}},\ \bibinfo
  {pages} {26} (\bibinfo {year} {1998})},\ \Eprint
  {http://arxiv.org/abs/astro-ph/9610120} {arXiv:astro-ph/9610120 [astro-ph]}
  \BibitemShut {NoStop}%
\bibitem [{\citenamefont {LoVerde}\ and\ \citenamefont
  {Afshordi}(2008)}]{LoVerde:2008re}%
  \BibitemOpen
  \bibfield  {author} {\bibinfo {author} {\bibfnamefont {M.}~\bibnamefont
  {LoVerde}}\ and\ \bibinfo {author} {\bibfnamefont {N.}~\bibnamefont
  {Afshordi}},\ }\href {\doibase 10.1103/PhysRevD.78.123506} {\bibfield
  {journal} {\bibinfo  {journal} {Phys. Rev.}\ }\textbf {\bibinfo {volume}
  {D78}},\ \bibinfo {pages} {123506} (\bibinfo {year} {2008})},\ \Eprint
  {http://arxiv.org/abs/0809.5112} {arXiv:0809.5112 [astro-ph]} \BibitemShut
  {NoStop}%
\bibitem [{\citenamefont {Lewis}\ \emph {et~al.}(2000)\citenamefont {Lewis},
  \citenamefont {Challinor},\ and\ \citenamefont {Lasenby}}]{Lewis:1999bs}%
  \BibitemOpen
  \bibfield  {author} {\bibinfo {author} {\bibfnamefont {A.}~\bibnamefont
  {Lewis}}, \bibinfo {author} {\bibfnamefont {A.}~\bibnamefont {Challinor}}, \
  and\ \bibinfo {author} {\bibfnamefont {A.}~\bibnamefont {Lasenby}},\ }\href
  {\doibase 10.1086/309179} {\bibfield  {journal} {\bibinfo  {journal}
  {Astrophys. J.}\ }\textbf {\bibinfo {volume} {538}},\ \bibinfo {pages} {473}
  (\bibinfo {year} {2000})},\ \Eprint {http://arxiv.org/abs/astro-ph/9911177}
  {arXiv:astro-ph/9911177 [astro-ph]} \BibitemShut {NoStop}%
\bibitem [{\citenamefont {{Blas}}\ \emph {et~al.}(2011)\citenamefont {{Blas}},
  \citenamefont {{Lesgourgues}},\ and\ \citenamefont
  {{Tram}}}]{2011JCAP...07..034B}%
  \BibitemOpen
  \bibfield  {author} {\bibinfo {author} {\bibfnamefont {D.}~\bibnamefont
  {{Blas}}}, \bibinfo {author} {\bibfnamefont {J.}~\bibnamefont
  {{Lesgourgues}}}, \ and\ \bibinfo {author} {\bibfnamefont {T.}~\bibnamefont
  {{Tram}}},\ }\href {\doibase 10.1088/1475-7516/2011/07/034} {\bibfield
  {journal} {\bibinfo  {journal} {\jcap}\ }\textbf {\bibinfo {volume} {7}},\
  \bibinfo {eid} {034} (\bibinfo {year} {2011})},\ \Eprint
  {http://arxiv.org/abs/1104.2933} {arXiv:1104.2933} \BibitemShut {NoStop}%
\bibitem [{\citenamefont {Takahashi}\ \emph {et~al.}(2012)\citenamefont
  {Takahashi}, \citenamefont {Sato}, \citenamefont {Nishimichi}, \citenamefont
  {Taruya},\ and\ \citenamefont {Oguri}}]{Takahashi:2012em}%
  \BibitemOpen
  \bibfield  {author} {\bibinfo {author} {\bibfnamefont {R.}~\bibnamefont
  {Takahashi}}, \bibinfo {author} {\bibfnamefont {M.}~\bibnamefont {Sato}},
  \bibinfo {author} {\bibfnamefont {T.}~\bibnamefont {Nishimichi}}, \bibinfo
  {author} {\bibfnamefont {A.}~\bibnamefont {Taruya}}, \ and\ \bibinfo {author}
  {\bibfnamefont {M.}~\bibnamefont {Oguri}},\ }\href {\doibase
  10.1088/0004-637X/761/2/152} {\bibfield  {journal} {\bibinfo  {journal}
  {\apj}\ }\textbf {\bibinfo {volume} {761}},\ \bibinfo {pages} {152} (\bibinfo
  {year} {2012})},\ \Eprint {http://arxiv.org/abs/1208.2701} {arXiv:1208.2701
  [astro-ph.CO]} \BibitemShut {NoStop}%
\bibitem [{\citenamefont {Mead}\ \emph {et~al.}(2015)\citenamefont {Mead},
  \citenamefont {Peacock}, \citenamefont {Heymans}, \citenamefont {Joudaki},\
  and\ \citenamefont {Heavens}}]{Mead:2015yca}%
  \BibitemOpen
  \bibfield  {author} {\bibinfo {author} {\bibfnamefont {A.}~\bibnamefont
  {Mead}}, \bibinfo {author} {\bibfnamefont {J.}~\bibnamefont {Peacock}},
  \bibinfo {author} {\bibfnamefont {C.}~\bibnamefont {Heymans}}, \bibinfo
  {author} {\bibfnamefont {S.}~\bibnamefont {Joudaki}}, \ and\ \bibinfo
  {author} {\bibfnamefont {A.}~\bibnamefont {Heavens}},\ }\href {\doibase
  10.1093/mnras/stv2036} {\bibfield  {journal} {\bibinfo  {journal} {\mnras}\
  }\textbf {\bibinfo {volume} {454}},\ \bibinfo {pages} {1958} (\bibinfo {year}
  {2015})},\ \Eprint {http://arxiv.org/abs/1505.07833} {arXiv:1505.07833
  [astro-ph.CO]} \BibitemShut {NoStop}%
\bibitem [{\citenamefont {Mead}\ \emph {et~al.}(2016)\citenamefont {Mead},
  \citenamefont {Heymans}, \citenamefont {Lombriser}, \citenamefont {Peacock},
  \citenamefont {Steele},\ and\ \citenamefont {Winther}}]{Mead:2016zqy}%
  \BibitemOpen
  \bibfield  {author} {\bibinfo {author} {\bibfnamefont {A.}~\bibnamefont
  {Mead}}, \bibinfo {author} {\bibfnamefont {C.}~\bibnamefont {Heymans}},
  \bibinfo {author} {\bibfnamefont {L.}~\bibnamefont {Lombriser}}, \bibinfo
  {author} {\bibfnamefont {J.}~\bibnamefont {Peacock}}, \bibinfo {author}
  {\bibfnamefont {O.}~\bibnamefont {Steele}}, \ and\ \bibinfo {author}
  {\bibfnamefont {H.}~\bibnamefont {Winther}},\ }\href {\doibase
  10.1093/mnras/stw681} {\bibfield  {journal} {\bibinfo  {journal} {\mnras}\
  }\textbf {\bibinfo {volume} {459}},\ \bibinfo {pages} {1468} (\bibinfo {year}
  {2016})},\ \Eprint {http://arxiv.org/abs/1602.02154} {arXiv:1602.02154
  [astro-ph.CO]} \BibitemShut {NoStop}%
\bibitem [{\citenamefont {Smith}\ \emph {et~al.}(2003)\citenamefont {Smith},
  \citenamefont {Peacock}, \citenamefont {Jenkins}, \citenamefont {White},
  \citenamefont {Frenk}, \citenamefont {Pearce}, \citenamefont {Thomas},
  \citenamefont {Efstathiou},\ and\ \citenamefont {Couchmann}}]{Smith:2002dz}%
  \BibitemOpen
  \bibfield  {author} {\bibinfo {author} {\bibfnamefont {R.~E.}\ \bibnamefont
  {Smith}}, \bibinfo {author} {\bibfnamefont {J.~A.}\ \bibnamefont {Peacock}},
  \bibinfo {author} {\bibfnamefont {A.}~\bibnamefont {Jenkins}}, \bibinfo
  {author} {\bibfnamefont {S.~D.~M.}\ \bibnamefont {White}}, \bibinfo {author}
  {\bibfnamefont {C.~S.}\ \bibnamefont {Frenk}}, \bibinfo {author}
  {\bibfnamefont {F.~R.}\ \bibnamefont {Pearce}}, \bibinfo {author}
  {\bibfnamefont {P.~A.}\ \bibnamefont {Thomas}}, \bibinfo {author}
  {\bibfnamefont {G.}~\bibnamefont {Efstathiou}}, \ and\ \bibinfo {author}
  {\bibfnamefont {H.~M.~P.}\ \bibnamefont {Couchmann}} (\bibinfo
  {collaboration} {VIRGO Consortium}),\ }\href {\doibase
  10.1046/j.1365-8711.2003.06503.x} {\bibfield  {journal} {\bibinfo  {journal}
  {\mnras}\ }\textbf {\bibinfo {volume} {341}},\ \bibinfo {pages} {1311}
  (\bibinfo {year} {2003})},\ \Eprint {http://arxiv.org/abs/astro-ph/0207664}
  {arXiv:astro-ph/0207664 [astro-ph]} \BibitemShut {NoStop}%
\bibitem [{\citenamefont {{Smith}}\ \emph {et~al.}(2018)\citenamefont
  {{Smith}}, \citenamefont {{Madhavacheril}}, \citenamefont {{M{\"u}nchmeyer}},
  \citenamefont {{Ferraro}}, \citenamefont {{Giri}},\ and\ \citenamefont
  {{Johnson}}}]{2018KSetallkSZ}%
  \BibitemOpen
  \bibfield  {author} {\bibinfo {author} {\bibfnamefont {K.~M.}\ \bibnamefont
  {{Smith}}}, \bibinfo {author} {\bibfnamefont {M.~S.}\ \bibnamefont
  {{Madhavacheril}}}, \bibinfo {author} {\bibfnamefont {M.}~\bibnamefont
  {{M{\"u}nchmeyer}}}, \bibinfo {author} {\bibfnamefont {S.}~\bibnamefont
  {{Ferraro}}}, \bibinfo {author} {\bibfnamefont {U.}~\bibnamefont {{Giri}}}, \
  and\ \bibinfo {author} {\bibfnamefont {M.~C.}\ \bibnamefont {{Johnson}}},\
  }\href@noop {} {\bibfield  {journal} {\bibinfo  {journal} {ArXiv e-prints}\ }
  (\bibinfo {year} {2018})},\ \Eprint {http://arxiv.org/abs/1810.13423}
  {arXiv:1810.13423} \BibitemShut {NoStop}%
\bibitem [{\citenamefont {Deutsch}\ \emph {et~al.}(2017)\citenamefont
  {Deutsch}, \citenamefont {Dimastrogiovanni}, \citenamefont {Johnson},
  \citenamefont {M{\"u}nchmeyer},\ and\ \citenamefont
  {Terrana}}]{Deutsch:2017ybc}%
  \BibitemOpen
  \bibfield  {author} {\bibinfo {author} {\bibfnamefont {A.-S.}\ \bibnamefont
  {Deutsch}}, \bibinfo {author} {\bibfnamefont {E.}~\bibnamefont
  {Dimastrogiovanni}}, \bibinfo {author} {\bibfnamefont {M.~C.}\ \bibnamefont
  {Johnson}}, \bibinfo {author} {\bibfnamefont {M.}~\bibnamefont
  {M{\"u}nchmeyer}}, \ and\ \bibinfo {author} {\bibfnamefont {A.}~\bibnamefont
  {Terrana}},\ }\href@noop {} {\  (\bibinfo {year} {2017})},\ \Eprint
  {http://arxiv.org/abs/1707.08129} {arXiv:1707.08129 [astro-ph.CO]}
  \BibitemShut {NoStop}%
\bibitem [{\citenamefont {Smith}\ \emph {et~al.}(2018)\citenamefont {Smith},
  \citenamefont {Madhavacheril}, \citenamefont {M{\"u}nchmeyer}, \citenamefont
  {Ferraro}, \citenamefont {Giri},\ and\ \citenamefont
  {Johnson}}]{Smith:2018bpn}%
  \BibitemOpen
  \bibfield  {author} {\bibinfo {author} {\bibfnamefont {K.~M.}\ \bibnamefont
  {Smith}}, \bibinfo {author} {\bibfnamefont {M.~S.}\ \bibnamefont
  {Madhavacheril}}, \bibinfo {author} {\bibfnamefont {M.}~\bibnamefont
  {M{\"u}nchmeyer}}, \bibinfo {author} {\bibfnamefont {S.}~\bibnamefont
  {Ferraro}}, \bibinfo {author} {\bibfnamefont {U.}~\bibnamefont {Giri}}, \
  and\ \bibinfo {author} {\bibfnamefont {M.~C.}\ \bibnamefont {Johnson}},\
  }\href@noop {} {\  (\bibinfo {year} {2018})},\ \Eprint
  {http://arxiv.org/abs/1810.13423} {arXiv:1810.13423 [astro-ph.CO]}
  \BibitemShut {NoStop}%
\bibitem [{\citenamefont {Green}\ \emph {et~al.}(2017)\citenamefont {Green},
  \citenamefont {Meyers},\ and\ \citenamefont {van Engelen}}]{Green:2016cjr}%
  \BibitemOpen
  \bibfield  {author} {\bibinfo {author} {\bibfnamefont {D.}~\bibnamefont
  {Green}}, \bibinfo {author} {\bibfnamefont {J.}~\bibnamefont {Meyers}}, \
  and\ \bibinfo {author} {\bibfnamefont {A.}~\bibnamefont {van Engelen}},\
  }\href {\doibase 10.1088/1475-7516/2017/12/005} {\bibfield  {journal}
  {\bibinfo  {journal} {JCAP}\ }\textbf {\bibinfo {volume} {1712}},\ \bibinfo
  {pages} {005} (\bibinfo {year} {2017})},\ \Eprint
  {http://arxiv.org/abs/1609.08143} {arXiv:1609.08143 [astro-ph.CO]}
  \BibitemShut {NoStop}%
\bibitem [{\citenamefont {Foreman}\ \emph {et~al.}(2018)\citenamefont
  {Foreman}, \citenamefont {Meerburg}, \citenamefont {Meyers},\ and\
  \citenamefont {van Engelen}}]{Foreman:2018lci}%
  \BibitemOpen
  \bibfield  {author} {\bibinfo {author} {\bibfnamefont {S.}~\bibnamefont
  {Foreman}}, \bibinfo {author} {\bibfnamefont {P.~D.}\ \bibnamefont
  {Meerburg}}, \bibinfo {author} {\bibfnamefont {J.}~\bibnamefont {Meyers}}, \
  and\ \bibinfo {author} {\bibfnamefont {A.}~\bibnamefont {van Engelen}},\
  }\href@noop {} {\  (\bibinfo {year} {2018})},\ \Eprint
  {http://arxiv.org/abs/1811.00529} {arXiv:1811.00529 [astro-ph.CO]}
  \BibitemShut {NoStop}%
\bibitem [{\citenamefont {Hirata}\ and\ \citenamefont
  {Seljak}(2003)}]{Hirata:2002jy}%
  \BibitemOpen
  \bibfield  {author} {\bibinfo {author} {\bibfnamefont {C.~M.}\ \bibnamefont
  {Hirata}}\ and\ \bibinfo {author} {\bibfnamefont {U.}~\bibnamefont
  {Seljak}},\ }\href {\doibase 10.1103/PhysRevD.67.043001} {\bibfield
  {journal} {\bibinfo  {journal} {Phys. Rev.}\ }\textbf {\bibinfo {volume}
  {D67}},\ \bibinfo {pages} {043001} (\bibinfo {year} {2003})},\ \Eprint
  {http://arxiv.org/abs/astro-ph/0209489} {arXiv:astro-ph/0209489 [astro-ph]}
  \BibitemShut {NoStop}%
\bibitem [{\citenamefont {Millea}\ \emph {et~al.}(2017)\citenamefont {Millea},
  \citenamefont {Anderes}, \citenamefont {Wandelt},\ and\ \citenamefont
  {Millea}}]{Millea:2017fyd}%
  \BibitemOpen
  \bibfield  {author} {\bibinfo {author} {\bibfnamefont {M.}~\bibnamefont
  {Millea}}, \bibinfo {author} {\bibfnamefont {E.}~\bibnamefont {Anderes}},
  \bibinfo {author} {\bibfnamefont {B.~D.}\ \bibnamefont {Wandelt}}, \ and\
  \bibinfo {author} {\bibfnamefont {M.}~\bibnamefont {Millea}},\ }\href@noop {}
  {\  (\bibinfo {year} {2017})},\ \Eprint {http://arxiv.org/abs/1708.06753}
  {arXiv:1708.06753 [astro-ph.CO]} \BibitemShut {NoStop}%
\bibitem [{\citenamefont {{Namikawa}}\ \emph {et~al.}(2013)\citenamefont
  {{Namikawa}}, \citenamefont {{Hanson}},\ and\ \citenamefont
  {{Takahashi}}}]{namikawa2013bias}%
  \BibitemOpen
  \bibfield  {author} {\bibinfo {author} {\bibfnamefont {T.}~\bibnamefont
  {{Namikawa}}}, \bibinfo {author} {\bibfnamefont {D.}~\bibnamefont
  {{Hanson}}}, \ and\ \bibinfo {author} {\bibfnamefont {R.}~\bibnamefont
  {{Takahashi}}},\ }\href {\doibase 10.1093/mnras/stt195} {\bibfield  {journal}
  {\bibinfo  {journal} {\mnras}\ }\textbf {\bibinfo {volume} {431}},\ \bibinfo
  {pages} {609} (\bibinfo {year} {2013})},\ \Eprint
  {http://arxiv.org/abs/1209.0091} {arXiv:1209.0091 [astro-ph.CO]} \BibitemShut
  {NoStop}%
\bibitem [{\citenamefont {{Zhang}}(2010)}]{Zhang10d}%
  \BibitemOpen
  \bibfield  {author} {\bibinfo {author} {\bibfnamefont {P.}~\bibnamefont
  {{Zhang}}},\ }\href {\doibase 10.1111/j.1745-3933.2010.00899.x} {\bibfield
  {journal} {\bibinfo  {journal} {MNRAS}\ }\textbf {\bibinfo {volume} {407}},\
  \bibinfo {pages} {L36} (\bibinfo {year} {2010})},\ \Eprint
  {http://arxiv.org/abs/1004.0990} {arXiv:1004.0990 [astro-ph.CO]} \BibitemShut
  {NoStop}%
\bibitem [{\citenamefont {Kamionkowski}\ and\ \citenamefont
  {Loeb}(1997)}]{Kamionkowski:1997na}%
  \BibitemOpen
  \bibfield  {author} {\bibinfo {author} {\bibfnamefont {M.}~\bibnamefont
  {Kamionkowski}}\ and\ \bibinfo {author} {\bibfnamefont {A.}~\bibnamefont
  {Loeb}},\ }\href {\doibase 10.1103/PhysRevD.56.4511} {\bibfield  {journal}
  {\bibinfo  {journal} {Phys. Rev.}\ }\textbf {\bibinfo {volume} {D56}},\
  \bibinfo {pages} {4511} (\bibinfo {year} {1997})},\ \Eprint
  {http://arxiv.org/abs/astro-ph/9703118} {arXiv:astro-ph/9703118 [astro-ph]}
  \BibitemShut {NoStop}%
\bibitem [{\citenamefont {Bunn}(2006)}]{Bunn:2006mp}%
  \BibitemOpen
  \bibfield  {author} {\bibinfo {author} {\bibfnamefont {E.~F.}\ \bibnamefont
  {Bunn}},\ }\href {\doibase 10.1103/PhysRevD.73.123517} {\bibfield  {journal}
  {\bibinfo  {journal} {Phys. Rev.}\ }\textbf {\bibinfo {volume} {D73}},\
  \bibinfo {pages} {123517} (\bibinfo {year} {2006})},\ \Eprint
  {http://arxiv.org/abs/astro-ph/0603271} {arXiv:astro-ph/0603271 [astro-ph]}
  \BibitemShut {NoStop}%
\bibitem [{\citenamefont {Deutsch}\ \emph {et~al.}(2018)\citenamefont
  {Deutsch}, \citenamefont {Johnson}, \citenamefont {M{\"u}nchmeyer},\ and\
  \citenamefont {Terrana}}]{Deutsch:2017cja}%
  \BibitemOpen
  \bibfield  {author} {\bibinfo {author} {\bibfnamefont {A.-S.}\ \bibnamefont
  {Deutsch}}, \bibinfo {author} {\bibfnamefont {M.~C.}\ \bibnamefont
  {Johnson}}, \bibinfo {author} {\bibfnamefont {M.}~\bibnamefont
  {M{\"u}nchmeyer}}, \ and\ \bibinfo {author} {\bibfnamefont {A.}~\bibnamefont
  {Terrana}},\ }\href {\doibase 10.1088/1475-7516/2018/04/034} {\bibfield
  {journal} {\bibinfo  {journal} {JCAP}\ }\textbf {\bibinfo {volume} {1804}},\
  \bibinfo {pages} {034} (\bibinfo {year} {2018})},\ \Eprint
  {http://arxiv.org/abs/1705.08907} {arXiv:1705.08907 [astro-ph.CO]}
  \BibitemShut {NoStop}%
\bibitem [{\citenamefont {Meyers}\ \emph {et~al.}(2018)\citenamefont {Meyers},
  \citenamefont {Meerburg}, \citenamefont {van Engelen},\ and\ \citenamefont
  {Battaglia}}]{Meyers:2017rtf}%
  \BibitemOpen
  \bibfield  {author} {\bibinfo {author} {\bibfnamefont {J.}~\bibnamefont
  {Meyers}}, \bibinfo {author} {\bibfnamefont {P.~D.}\ \bibnamefont
  {Meerburg}}, \bibinfo {author} {\bibfnamefont {A.}~\bibnamefont {van
  Engelen}}, \ and\ \bibinfo {author} {\bibfnamefont {N.}~\bibnamefont
  {Battaglia}},\ }\href {\doibase 10.1103/PhysRevD.97.103505} {\bibfield
  {journal} {\bibinfo  {journal} {Phys. Rev.}\ }\textbf {\bibinfo {volume}
  {D97}},\ \bibinfo {pages} {103505} (\bibinfo {year} {2018})},\ \Eprint
  {http://arxiv.org/abs/1710.01708} {arXiv:1710.01708 [astro-ph.CO]}
  \BibitemShut {NoStop}%
\bibitem [{\citenamefont {Seljak}(2009)}]{Seljak:2008xr}%
  \BibitemOpen
  \bibfield  {author} {\bibinfo {author} {\bibfnamefont {U.}~\bibnamefont
  {Seljak}},\ }\href {\doibase 10.1103/PhysRevLett.102.021302} {\bibfield
  {journal} {\bibinfo  {journal} {Phys. Rev. Lett.}\ }\textbf {\bibinfo
  {volume} {102}},\ \bibinfo {pages} {021302} (\bibinfo {year} {2009})},\
  \Eprint {http://arxiv.org/abs/0807.1770} {arXiv:0807.1770 [astro-ph]}
  \BibitemShut {NoStop}%
\bibitem [{\citenamefont {Dalal}\ \emph {et~al.}(2008)\citenamefont {Dalal},
  \citenamefont {Dore}, \citenamefont {Huterer},\ and\ \citenamefont
  {Shirokov}}]{Dalal:2007cu}%
  \BibitemOpen
  \bibfield  {author} {\bibinfo {author} {\bibfnamefont {N.}~\bibnamefont
  {Dalal}}, \bibinfo {author} {\bibfnamefont {O.}~\bibnamefont {Dore}},
  \bibinfo {author} {\bibfnamefont {D.}~\bibnamefont {Huterer}}, \ and\
  \bibinfo {author} {\bibfnamefont {A.}~\bibnamefont {Shirokov}},\ }\href
  {\doibase 10.1103/PhysRevD.77.123514} {\bibfield  {journal} {\bibinfo
  {journal} {Phys. Rev.}\ }\textbf {\bibinfo {volume} {D77}},\ \bibinfo {pages}
  {123514} (\bibinfo {year} {2008})},\ \Eprint {http://arxiv.org/abs/0710.4560}
  {arXiv:0710.4560 [astro-ph]} \BibitemShut {NoStop}%
\bibitem [{\citenamefont {Schmittfull}\ and\ \citenamefont
  {Seljak}(2018)}]{Schmittfull:2017ffw}%
  \BibitemOpen
  \bibfield  {author} {\bibinfo {author} {\bibfnamefont {M.}~\bibnamefont
  {Schmittfull}}\ and\ \bibinfo {author} {\bibfnamefont {U.}~\bibnamefont
  {Seljak}},\ }\href {\doibase 10.1103/PhysRevD.97.123540} {\bibfield
  {journal} {\bibinfo  {journal} {Phys. Rev.}\ }\textbf {\bibinfo {volume}
  {D97}},\ \bibinfo {pages} {123540} (\bibinfo {year} {2018})},\ \Eprint
  {http://arxiv.org/abs/1710.09465} {arXiv:1710.09465 [astro-ph.CO]}
  \BibitemShut {NoStop}%
\bibitem [{\citenamefont {M{\"u}nchmeyer}\ \emph {et~al.}(2018)\citenamefont
  {M{\"u}nchmeyer}, \citenamefont {Madhavacheril}, \citenamefont {Ferraro},
  \citenamefont {Johnson},\ and\ \citenamefont {Smith}}]{Munchmeyer:2018eey}%
  \BibitemOpen
  \bibfield  {author} {\bibinfo {author} {\bibfnamefont {M.}~\bibnamefont
  {M{\"u}nchmeyer}}, \bibinfo {author} {\bibfnamefont {M.~S.}\ \bibnamefont
  {Madhavacheril}}, \bibinfo {author} {\bibfnamefont {S.}~\bibnamefont
  {Ferraro}}, \bibinfo {author} {\bibfnamefont {M.~C.}\ \bibnamefont
  {Johnson}}, \ and\ \bibinfo {author} {\bibfnamefont {K.~M.}\ \bibnamefont
  {Smith}},\ }\href@noop {} {\  (\bibinfo {year} {2018})},\ \Eprint
  {http://arxiv.org/abs/1810.13424} {arXiv:1810.13424 [astro-ph.CO]}
  \BibitemShut {NoStop}%
\bibitem [{\citenamefont {Linder}(2005)}]{Linder:2005in}%
  \BibitemOpen
  \bibfield  {author} {\bibinfo {author} {\bibfnamefont {E.~V.}\ \bibnamefont
  {Linder}},\ }\href {\doibase 10.1103/PhysRevD.72.043529} {\bibfield
  {journal} {\bibinfo  {journal} {Phys. Rev.}\ }\textbf {\bibinfo {volume}
  {D72}},\ \bibinfo {pages} {043529} (\bibinfo {year} {2005})},\ \Eprint
  {http://arxiv.org/abs/astro-ph/0507263} {arXiv:astro-ph/0507263 [astro-ph]}
  \BibitemShut {NoStop}%
\bibitem [{\citenamefont {Linder}\ and\ \citenamefont
  {Cahn}(2007)}]{Linder:2007hg}%
  \BibitemOpen
  \bibfield  {author} {\bibinfo {author} {\bibfnamefont {E.~V.}\ \bibnamefont
  {Linder}}\ and\ \bibinfo {author} {\bibfnamefont {R.~N.}\ \bibnamefont
  {Cahn}},\ }\href {\doibase 10.1016/j.astropartphys.2007.09.003} {\bibfield
  {journal} {\bibinfo  {journal} {Astropart. Phys.}\ }\textbf {\bibinfo
  {volume} {28}},\ \bibinfo {pages} {481} (\bibinfo {year} {2007})},\ \Eprint
  {http://arxiv.org/abs/astro-ph/0701317} {arXiv:astro-ph/0701317 [astro-ph]}
  \BibitemShut {NoStop}%
\bibitem [{\citenamefont {Bond}\ and\ \citenamefont
  {Szalay}(1983)}]{Bond:1983hb}%
  \BibitemOpen
  \bibfield  {author} {\bibinfo {author} {\bibfnamefont {J.~R.}\ \bibnamefont
  {Bond}}\ and\ \bibinfo {author} {\bibfnamefont {A.~S.}\ \bibnamefont
  {Szalay}},\ }\href {\doibase 10.1086/161460} {\bibfield  {journal} {\bibinfo
  {journal} {Astrophys. J.}\ }\textbf {\bibinfo {volume} {274}},\ \bibinfo
  {pages} {443} (\bibinfo {year} {1983})}\BibitemShut {NoStop}%
\end{thebibliography}%

\end{document}